\documentclass[aps,prb,twocolumn,superscriptaddress,showpacs]{revtex4-1}
\usepackage{graphicx}% Include figure files

\begin{document}
\newcommand{\ket}[1]{\left|#1\right\rangle}
\newcommand{\bra}[1]{\left\langle#1\right|}
\newcommand{\im}{i}
\newcommand{\imag}{\grave{\imath}}
%\preprint{}

\title{Transmon-phonon coupling of plasma oscillations and lattice vibrations}

\author{A. J. Skinner}
%\email[]{skinner@wam.umd.edu}
%\thanks{}
\affiliation{Welkin Research, Saratoga Springs, NY 12866}


\date{\today}

\begin{abstract}
In the transmon qubit we expect from conservation of momentum and energy a coupling between the plasma oscillations and the vibrations of the underlying lattice.  Specifically, the electron velocities and their kinetic energy density are boosted by the underlying lattice vibrations.  We consider this effect in a representative transmon comprising two semi-circular superconducting charge islands joined by a Josephson junction.  In particular, we solve the Fourier transform of a two-dimensional radial current density having inversion symmetry.  The resulting spectral density is ohmic but also scales quadratically with the critical current $I_c$ and logarithmically with the size of the transmon: $J(\omega)\sim I_c^2 \; \omega \log(kR)$.  We make positive-definite Born-Markov approximations in a generalized Fermi's Golden Rule and estimate the phonon-induced dephasing rate is negligible compared to current experiments.
\end{abstract}

%\pacs{03.65.Yz}
%\keywords{}

\maketitle

\section{Introduction}

The transmon, or transmission-line shunted plasma oscillation qubit, is a superconducting Josephson junction qubit situated in a microwave cavity.\cite{Koch,Schuster, Schreier} The junction connects a pair of superconducting islands whose charging energy $E_C$ is significantly smaller than the Josephson energy $E_J$.   The large ratio $E_J/E_C \gg 1$ reduces sensitivity to charge noise relative to the anharmonicity and thus improves dephasing times relative to the speed of qubit operation.  The use of a three-dimensional cavity \cite{Paik} and other refinements \cite{Houck2} have improved dephasing times to almost $100$ $\mu$s. \cite{Rigetti}

The transmon is similar to the Cooper-pair box (CPB) qubit \cite{Bouchiat,Nakamura} but with a significantly larger ratio $E_J/E_C$.  To achieve low charging energies the islands can be expansive, with dimensions measured in hundreds of microns and combined area approaching $0.5$ mm$^2$.  It is typically fabricated on a substrate such as sapphire, with the junction a thin oxidized layer such as Al$_2$O$_3$ between islands of superconducting Niobium or Aluminum.  It is then placed in an electromagnetic cavity where it is measured, controlled, and coupled to other transmons, by microwaves.\cite{Majer}

In the $E_J/E_C \gg 1$ regime the transmon is similar to an anharmonic oscillator (cosine potential) but with the number of excess electron pairs $n$ and the superconducting phase difference $\phi$ as the conjugate variables.  A superposition of the lowest eigenstates $\ket{0}$ and $\ket{1}$ will result in oscillations of the current $I_c \sin(\phi)$ in which the total momentum of the charge carriers is also oscillating.  Conservation of momentum suggests a recoil of the underlying lattice.  The relatively large area of the transmon also encompasses many lattice sites whose vibrations may interfere with qubit operation.

\begin{figure}
\begin{center}
\includegraphics{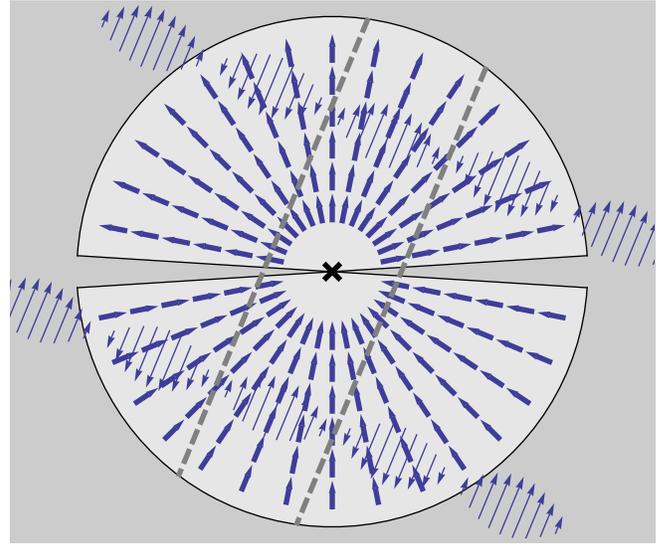}
\end{center}
\caption{Model transmon comprising two semi-circular superconducting islands joined by a Josephson junction (bold x).  Radial current density (thick arrows) flows off one island, through the junction, and onto the other.  A phonon's transverse in-plane motion (thin arrows) collectively boosts the electrons' kinetic energy, especially within a diametrical band (dashed borders) aligned with the transverse motion.}
\label{TransmonJandP}
\end{figure}

In the case of superconducting quantum interference device (SQUID) qubits, the coupling to phonons was previously derived from the boost that lattice vibrations give to the electron velocities and thus their kinetic energy density. \cite{Chudnovsky}  Here we apply the same technique to a representative transmon qubit using a radial model for the current density.  We solve its Fourier transform and find the spectral density is ohmic, but also quadratic in the critical current, and scales logarithmically with the size, in phonon wavelengths, of the transmon: $J(\omega)\sim I_c^2 \; \omega \log(kR)$.  We also use positive-definite Born-Markov approximations in a generalized Fermi's Golden Rule to estimate the phonon-induced dephasing time is orders of magnitude longer than state of the art experiments.

\section{Eigenstates and Energies}

The total Hamiltonian comprises that of the transmon, $H_S$, the phonon bath, $H_B$, and their coupling, $V$:

\begin{equation}
H=H_S+H_B+V.
\end{equation}

\subsection{The Transmon CPB Hamiltonian}

Neglecting the microwave cavity, the transmon's effective Hamilton can be reduced to CPB form \cite{Koch}
\begin{equation}
H_S = 4 E_C (n-n_g)^2 - E_J \cos \phi,
\label{H_S}
\end{equation}
with $\phi$ as the phase difference across the junction, $n = -\im \frac{d}{d\phi}$ as the number of excess Cooper-pairs on one island, and $n_g$ as the effective offset charge in pair-charge units $2e$.  

By introducing the function $g(x)\equiv e^{-2\im n_g x}\psi(2x)$ the eigenvalue equation becomes the Mathieu equation
\begin{equation}
g''(x)+\left[\underbrace{\frac{E}{E_C}}_{a}-2\underbrace{(\frac{-E_J}{2 E_C})}_{q}\right]g(x)=0
\end{equation}
whose solutions $me_{\nu}(q,x)$ can then be used to reconstruct
\begin{equation}
\psi(\phi) = \frac{1}{\sqrt{2\pi}} e^{\im n_g \phi} me_{\nu}(q,\phi/2).
\end{equation}

The boundary condition $\psi(\phi)=\psi(\phi+2\pi)$ restricts the Mathieu functions $me_{\nu}(q,x)$ to characteristic exponents $\nu=-2(n_g-n_c)$ with $n_c$ an integer number of extra cycles of $\psi(\phi)$ fitting into a full $2\pi$ period of $\phi$ .  The Mathieu characteristic eigenvalues $a(\nu,q)=E/E_C$ thus separate into bands with $n_c=n_c(m,n_g)$ sorting the eigenvalues according to band index $m$.

\begin{figure}
\begin{center}
\includegraphics{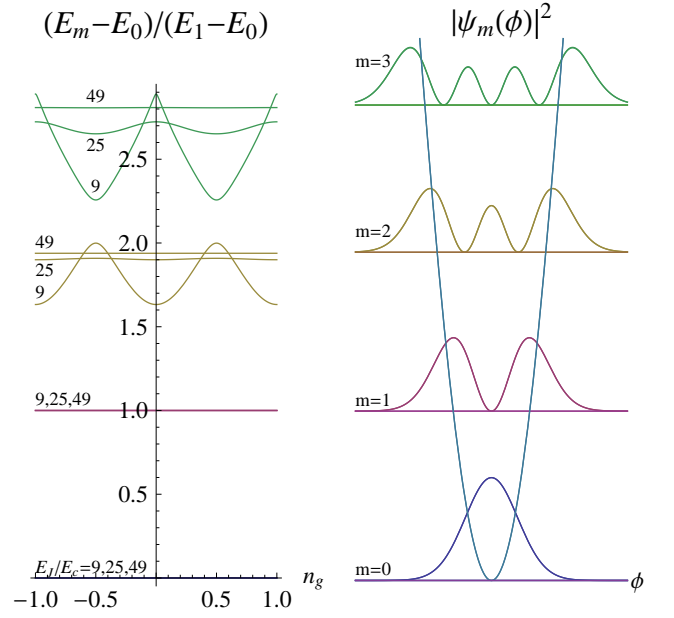}
\end{center}
\caption{Transmon energy levels and probability densities in a cosine potential.  Left plot: lowest four bands with charge dispersion and anharmonicity for $E_J/E_C=9,25,49$.  Right plot: probability densities for $E_J/E_C = 49$.}
\label{levelsandstates}
\end{figure}

As shown in Figure~\ref{levelsandstates}, the charge dispersions, or band widths, decrease exponentially with $E_J/E_C$ while the transmon's relative anharmonicity, $(E_{21}-E_{10})/E_{10}$, decreases only algebraically.\cite{Koch}  Reducing the anharmonicity prolongs the time needed for qubit operations but the exponentially reduced dispersion more than compensates with extended coherence times.  This is the advantage of the large $E_J/E_C$ ratio in the transmon qubit.

Within the $\{\ket{0}\equiv\ket{\psi_0},\ket{1}\equiv\ket{\psi_1}\}$ logical subspace we have $H_S \sim \sigma_z \equiv \ket{0}\bra{0}-\ket{1}\bra{1}$.  Since the current through the junction $I_c \sin \phi$ couples to the phonons we will soon also be using the operator $\sigma_{\phi}\equiv \sin\phi$.  We calculate its non-zero (off-diagonal) transition matrix elements by numerical integration:
\begin{equation}
\bra{0}\sigma_{\phi}\ket{1} = \int_{-\pi}^{\pi}\! d\phi \; \psi_0^*(\phi) \sin(\phi) \psi_1(\phi).
\end{equation}
For $E_J/E_C = 49$, as an example, $\bra{0}\sigma_{\phi}\ket{1} \approx 0.414 \im$ which is not far from the $1/\sqrt{2\alpha} \approx 0.450$ that one would get from a small angle approximation $\sin(\phi) \approx \phi$ in a not-anharmonic oscillator with $\alpha = \sqrt{(E_J/E_C)/8}$.

\subsection{The Harmonic Phonon Bath}
The harmonic crystal Hamiltonian is composed of phonon modes labelled by wavevector $\vec{k}$ and polarization index $s$.  Each phonon contributes an energy $\hbar \omega_{\vec{k}s}$:
\begin{equation}
H_B=\sum_{\vec{k},s}\hbar \omega_{\vec{k}s} a^{\dagger}_{\vec{k}s}a_{\vec{k}s}.
\label{H_B}
\end{equation}
Here $a^{\dagger}_{\vec{k}s}$ and $a_{\vec{k}s}$ are the phonon creation and annihilation operators.  We assume linear dispersions, $\omega_{1,2}(\vec{k}) = c_T\,  |\vec{k}|$ and $\omega_3(\vec{k}) = c_L\,  |\vec{k}|$, for transverse and longitudinal polarizations $\hat{e}_s(\vec{k})$.  The lattice site speeds are
\begin{equation}
\dot{\vec{u}}(\vec{r}) = \sum_{\vec{k}s} \sqrt{\frac{{\hbar \omega_{\vec{k}s}}}{2 \rho V}} \left(a_{\vec{k}s}\frac{e^{\im \vec{k}\cdot\vec{r}}}{\im}-a^{\dagger}_{\vec{k}s}\frac{e^{-\im \vec{k}\cdot\vec{r}}}{\im}\right) \hat{e}_s(\vec{k})
\label{momentumdensity}
\end{equation} with $\rho$ and $V$ the mass density and volume of the lattice.

With the direction of phonon propagation $\hat{k}$ specified by angles $\theta$ and $\phi$ measured, respectively, from $\hat{z}$ perpendicular to the transmon and $\hat{x}$ transecting it, and choosing $\hat{e}_1$ as the transverse in-plane polarization, the polarizations have $\{x,y,z\}$ components as follows:
\[
\begin{array}{rcccr}
\hat{e}_1\!=\! (\hat{k}\times\hat{z})/|\hat{k}\times\hat{z}| =  \{&\sin \phi& -\cos \phi& 0 &\} \\
\hat{e}_2\!=\! \hat{e}_3 \times \hat{e}_1 =  \{&\cos\phi\cos\theta & \sin\phi\cos\theta &-\sin\theta&\} \\
\hat{e}_3\!= \!\hat{k} =  \{&\cos\phi\sin\theta & \sin\phi\sin\theta & \cos\theta&\} \\
\end{array}
\]

\subsection{The Transmon-Phonon Coupling}
The coupling between the transmon and the crystal lattice arises from the fact that the current is formed from the electronic band states in the reference frame co-moving with the lattice sites during the vibrations of the crystal. \cite{Chudnovsky}

In the lab frame the electron velocity $\vec{v}_e$ must therefore include the speed $\dot{\vec{u}}$ of the lattice sites: $\vec{v}_e = \vec{j}/(e n_e)+\dot{\vec{u}}$.  Here $\vec{j}$ is the current density and $e$ and $n_e$ are the electron charge and number density of the electrons.  Their kinetic energy density $n_e \frac{1}{2}  m_e |\vec{v}_e|^2$ thus acquires a cross term and the transmon-phonon coupling $V$ accounts for this extra energy density integrated over the transmon:
\begin{equation}
V = \frac{m_e}{e}\int d^3r  \, \vec{j}\cdot\dot{\vec{u}}.
\label{kineticenergydensity}
\end{equation}

Qualitatively, the charge carriers converge radially inward from one semi-circular island, flow through the junction, and diverge radially outward onto the other island.  The junction current, $I_c \sigma_{\phi}$, flows at a radius $r$ through an area $\pi r T$ with $T$ as the thickness of the islands.  Our current density model is thus radial above the $x$-axis and opposite below and decreases like $1/\pi r T$ with distance $r$ from the junction:
\begin{equation}
\vec{j}(\vec{r}) =  \frac{I_c \sigma_{\phi}} {\pi r T}  \frac{y}{|y|}\hat{r}.
\end{equation}

Integrating $\vec{j}\cdot\dot{\vec{u}}$ over the transmon amounts to solving the components  $j_{\vec{k}s}$ of the Fourier transform of the current density.    In the thin-transmon approximation (thickness $T\ll\lambda/2\pi$ for relevant phonon wavelengths $\lambda\approx 1$ $\mu$m) we need only integrate in the $xy$-plane of the transmon out to its radius $R$.  In polar coordinates $(r,\psi)$, with $\psi$ measured from the $x$-axis transecting the transmon, we define a dimensionless version of this Fourier transform,
\begin{equation}
\tilde{j}_{\vec{k}s} \equiv \int_0^{kR} d(kr) \int_{-\pi}^{\pi} d\psi \frac{\psi}{|\psi|} \hat{r}\cdot\hat{e}_s(\vec{k}) e^{-\im k r \hat{k}\cdot\hat{r}}
\end{equation}
to write the coupling coefficients as
\begin{equation}
g_{\vec{k}s} \equiv \frac{1}{\im}\frac{m_e}{e}\sqrt{\frac{{\hbar \omega_{\vec{k}s}}}{2 \rho V}} \frac{I_c}{\pi}\frac{1}{k} \; \tilde{j}_{\vec{k}s}
\end{equation}
in terms of which the transmon-phonon coupling is
\begin{equation}
V = \sigma_{\phi} \otimes \sum_{\vec{k}s} \left(g_{\vec{k}s} a_{\vec{k}s}+g_{\vec{k}s}^*a^{\dagger}_{\vec{k}s}\right).
\label{kineticenergydensity}
\end{equation}

\section{The Fourier Transform}

\subsection{Transverse in-plane polarization}

Transverse in-plane phonons,  as shown for example in Figure~\ref{TransmonJandP}, are the relevant polarization for boosting the transmon's electron velocities.  Whatever their direction of propagation, there is always a diametrical band within which the current flow is qualitatively aligned with the phonon's transverse in-plane collective motion.

\begin{figure}
\begin{center}
\includegraphics{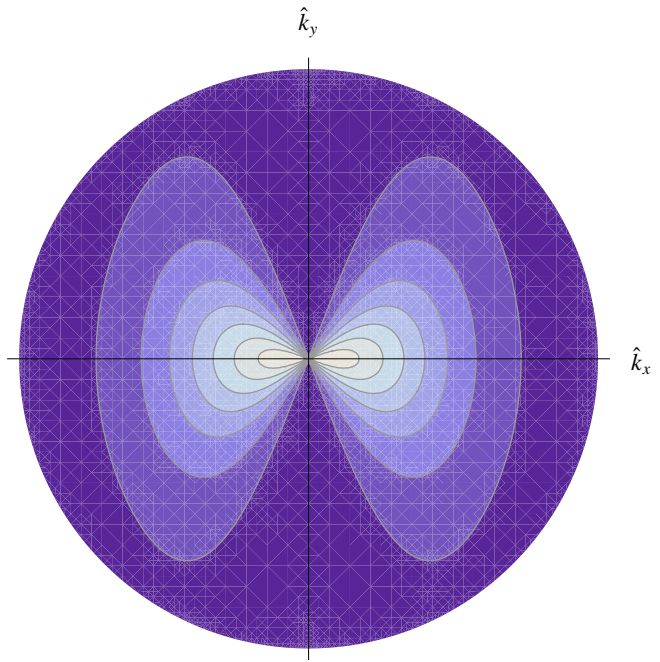}
\end{center}
\caption{Directional emission pattern $|j_{\vec{k}1}|^2$ of transverse in-plane phonons at $kR=5000$.}
\label{PolarKCap}
\end{figure}

Quantitatively, we take advantage of the inversion symmetry to integrate $\cos(kr \hat{k}\cdot\hat{r}) \hat{r}\cdot\hat{e}_1$ twice over one island.  With $\hat{k}\cdot\hat{r} = \cos(\phi-\psi)\sin\theta$ and $\hat{r}\cdot\hat{e}_1=\sin(\phi-\psi)$ we find
\begin{equation}
\tilde{j}_{\vec{k}1} = -4 \frac{\mbox{Si}[kR \cos\phi\sin\theta]}{\sin\theta} = -4 \frac{\mbox{Si}[k_x R]}{\sin\theta}
\end{equation}
where Si[ ] is the sine integral function.

The corresponding directional emission pattern for transverse in-plane phonons, $|j_{\vec{k}1}|^2 = 4^2 \mbox{Si}^2[k_xR]/\sin^2\theta$, is shown in Figure~\ref{PolarKCap}.  Steeply incident ($\theta\ll1$) phonons dominate the pattern because they have relatively long in-plane wavelengths $\lambda/\sin\theta$ that give a wider band within which the coupling accumulates.  On the other hand, phonons propagating in the $\pm \hat{y}$ directions have their band of transverse motion straddling the islands' gap and the in- and out-flows cancel.  Accordingly, the phonon-induced decoherence is mostly effected by steeply incident phonons in the $x,z$ plane.

Solving the spectral density will involve integration over all directions in $k$-space.  We obtain by numerical integration, for a variety of $250<kR<7500$, a good fit
\begin{equation}
\int d^2\Omega \;  |\tilde{j}_{\vec{k}1}|^2 = \int_0^{\pi} \! d\theta \sin\theta \int_0^{2\pi}\!d\phi \;  |\tilde{j}_{\vec{k}1}|^2 \approx a \log[b \; kR]
\end{equation}
with $a \approx 496$ and $b \approx 1.16$.

\section{Positive-Definite Born-Markov Approximations}

In this section we give an overview of techniques developed by Taj et al.\ \cite{Taj} to generalize Fermi's Golden Rule to nanodevices with positive-definite Born-Markov approximations.

The unperturbed Hamiltonian $H_0 \equiv H_S + H_B$ defines a standard interaction picture $U_t=e^{-\im H_0 t/\hbar}$ in which the coupling $\tilde{V}(t) = U_t^{\dagger} V U_t $ determines the evolution of the system-bath density operator $\tilde{\rho}(t)=U_t^{\dagger} \rho(t) U_t$ as
\begin{equation}
\Delta\tilde{\rho} = -\frac{1}{\hbar^2} \int_{t-\Delta t/2}^{t+\Delta t/2} dt_1 \int_{0}^{t_1} dt_2 [\tilde{V}(t_1),[\tilde{V}(t_2),\tilde{\rho}(t_2)]], 
\end{equation}
where we have neglected a first order term whose contribution to the system density operator $\tilde{\rho}_S$ will vanish, for a thermal bath, in the averaging $\Delta \tilde{\rho}_S = \mbox{Tr}_B[\Delta \tilde{\rho}]$ over bath eigenstates.  

\begin{figure}
\begin{center}
\includegraphics{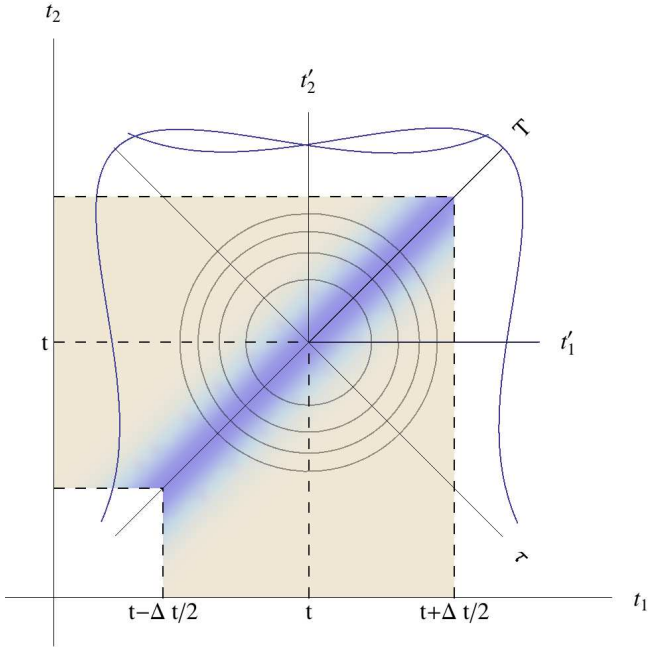}
\end{center}
\caption{$\Delta \tilde{\rho}$ arises from $[\tilde{V}(t_1),[\tilde{V}(t_2),\tilde{\rho}(t_2)]]$ (caricature dark shading) integrated within the upright trapezoid.  Symmetrizing the integral around $t_2=t_1$ discards self-adjoint shifts to $H_0$.  Positive-definite Born-Markov approximations are made with a Gaussian cutoff in the relative time $\tau$ and a commensurate Gaussian smoothing in the average time $T$. \cite{Taj}}
\label{t1t2plane}
\end{figure}

Simply dividing $\Delta \tilde{\rho}$ by $\Delta t$ gives a course-grained approximation to $\dot{\tilde{\rho}}$ ready for Born and Markov approximations.  However, the dynamics of $\rho$ can be kept physical (positive-definite) by an improved nested-integration around $t_1'\equiv t_1-t\approx 0$ and $t_2'\equiv t_2-t\approx 0$ using a Gaussian cutoff in the relative time $\tau\equiv t_1'-t_2'$ and a Gaussian smoothing in the ``macroscopic'' average time $T\equiv(t_1'+t_2')/2$ as follows:\cite{Davies,Taj}
\begin{widetext}
\begin{equation}
\dot{\tilde{\rho}} = -\frac{1}{\hbar^2} \int_{-\infty}^{\infty} dT \frac{e^{-T^2/2\bar{T}^2}}{\sqrt{2\pi}\bar{T}} \frac{1}{2}\int_ {-\infty} ^{\infty} d\tau e^{-\tau^2/2\bar{\tau}^2} [U_t^{\dagger}\tilde{V}(\overbrace{T+\tau/2}^{t_1'})U_t,[U_t^{\dagger}\tilde{V}(\overbrace{T-\tau/2}^{t_2'})U_t,\tilde{\rho}(t)]]
\end{equation}
\end{widetext}
Here we have made Born's approximation by replacing $\tilde{\rho}(t_2)$ with $\tilde{\rho}(t)$ and Markov's approximation by using a Gaussian cutoff time $\bar{\tau}$ longer than the correlation times for the weak coupling $\tilde{V}(t)$.  In extending the relative time integral back to $-\infty$, we have also disposed of the self-adjoint parts of the double commutator structure that could just as well have been included as small shifts to the unperturbed Hamiltonian $H_0$.

Finally, the smoothing time is chosen as $\bar{T}\equiv\bar{\tau}/2$ so that the Gaussian product re-factors as
\begin{equation}
e^{-T^2/2\bar{T}^2} e^{-\tau^2/2\bar{\tau}^2} =  e^{-t_1'^2/\bar{\tau}^2} e^{-t_2'^2/\bar{\tau}^2}
\end{equation}
in the $t_1',t_2'$ plane and the integrals separate in which case the evolution of the density operator may be written as
\begin{equation}
\dot{\tilde{\rho}}=-\frac{2\pi}{\hbar^2}\frac{1}{2} [U_t^{\dagger}LU_t,[U_t^{\dagger}LU_t,\tilde{\rho}]]
\end{equation}
in the interaction picture, or as
\begin{equation}
\dot {\rho} =-\frac{\im}{\hbar}[H_0,\rho] -\frac{2\pi}{\hbar^2}\frac{1}{2} [L,[L,\rho]]
\end{equation}
in the Schrodinger picture, with the Lindblad operator
\begin{equation}
L \equiv \sqrt{\bar{\omega}/\sqrt{2\pi^3}} \int_{-\infty}^{\infty} dt' \tilde{V}(t') e^{-\bar{\omega}^2t'^2}
\end{equation}
whose matrix elements between final and initial states are like the square root of a delta function as $\bar{\omega}\equiv 1/\bar{\tau} \rightarrow 0$:
\begin{equation}
\frac{2\pi}{\hbar^2}|\bra{f}L\ket{i}|^2 = \frac{2\pi}{\hbar^2}|\bra{f}V\ket{i}|^2 \underbrace{\left|\frac{e^{-(\omega_f-\omega_i)^2/4\bar{\omega}^2}}{\sqrt{\sqrt{2 \pi}\bar{\omega}}}\right|^2}_{\rightarrow \delta(\omega_f-\omega_i)}.
\end{equation}
We thus see that Fermi's original Golden Rule is embedded in the more general positive-definite double commutator structure of $[L,[L,\rho]]$.

\section{Transmon Dephasing Rate $\Gamma_{10}$}

\subsection{Master Equation}
The master equation for the transmon system is obtained by tracing over the bath,
\begin{equation}
\dot{\rho}_S = -\frac{\im}{\hbar} [H_S,\rho_S] -\frac{2\pi}{\hbar^2}\frac{1}{2} \mbox{Tr}_B\left[L,[L,\rho_S\otimes\rho_B]\right],
\end{equation}
where we start with the transmon $\rho_S$ initially uncorrelated with a thermal bath $\rho_B \equiv (e^{-H_B/k_BT})/Z_B$ and extend Born's approximation by assuming the (small) changes in the interaction picture $\tilde{\rho}$ occur mainly in the transmon: $\tilde{\rho}(t) =  \tilde{\rho}_S(t)\otimes \tilde{\rho}_B$ and thus $\rho(t)=\rho_S(t)\otimes\rho_B$.
 
Recalling $V=\sum_{\vec{k}s} \sigma_{\phi}\otimes (g_{\vec{k}s}a_{\vec{k}s}+g_{\vec{k}s}^*a_{\vec{k}s}^{\dagger})$ we solve in the two-dimensional logical eigenbasis of $H_S$ the Lindblad operators $L_{\vec{k}s}$ for each mode of the bath, i.e.\
\begin{equation}
L_{\vec{k}s} = 
\left[
\begin{array}{cc}
  0 &  \bra{0}\sigma_{\phi}\ket{1} g_{\vec{k}s}^* a_{\vec{k}s}^{\dagger}  \\
  \bra{1}\sigma_{\phi}\ket{0} g_{\vec{k}s} a_{\vec{k}s} & 0 
\end{array}
\right] \sqrt{\delta}
\end{equation}
where $\sqrt{\delta} \approx \sqrt{\delta(\omega_{\vec{k}s}-\omega)}$, in terms of the transmon frequency $\omega\equiv\omega_1-\omega_0$, and we have kept only the energy-conserving terms in the transitions between $\ket{0}$ and $\ket{1}$. \footnote{The raising and lowering bath operators that we discarded would have ultimately vanished, had we kept hold of them, in the subsequent trace over the sum over the bath modes.}  

Taking $L=\sum_{\vec{k}s} L_{\vec{k}s}$ and writing
\begin{equation}
\rho_S\otimes\rho_B =
\left[
\begin{array}{cc}
\rho_{00} &  \rho_{01}  \\
\rho_{10} & \rho_{11} 
 \end{array}
\right] \rho_B,
\end{equation}
with $\rho_{mn}\equiv\bra{m}\rho_S\ket{n}$, we multiply out the $2\times2$ matrices and trace the bath, making frequent use of the cyclic property of the trace and $\mbox{Tr}_B[a_{\vec{k}s}^{\dagger}a_{\vec{k}'s'}\rho_B]=\delta_{\vec{k}\vec{k}'}\delta_{ss'}N_{\omega_{\vec{k}s}}$ in terms of the thermal average occupation numbers $N_{\omega_{\vec{k}s}} \equiv 1/(e^{\hbar \omega_{\vec{k}s}/k_BT}-1)$.  We find
\begin{widetext}
\begin{equation}
\mbox{Tr}_B\!\left[L,[L,\rho]\right] \!=\! |\!\bra{0}\sigma_{\phi}\ket{1}\!|^2 \underbrace{\frac{V}{(2\pi)^3} \frac{1}{c_T}k^2 \int d^2\Omega \frac{m_e^2}{e^2} \frac{\hbar \omega}{2 \rho V} \frac{I_c^2}{\pi^2} \frac{1}{k^2} |\tilde{j}_{\vec{k}1}|^2}_{J_1(\omega) \equiv \sum_{\vec{k},s=1} |g_{\vec{k}s}|^2 \delta(\omega_{\vec{k}s}-\omega)} 
\left[
\begin{array}{cc}
 2 (N_{\omega} \rho_{00} -(N_{\omega}\!+1)\rho_{11}) & (2 N_{\omega}\!+1)\rho_{01} \\
(2N_{\omega}\!+1) \rho_{10} & 2 ((N_{\omega}\!+1)\rho_{11}-N_{\omega} \rho_{00} )
\end{array}
\right]
\end{equation}
\end{widetext}
where we have converted the $\vec{k}$-space sum to an integral, integrated over the delta function, and kept only the ($s=1$) in-plane polarization.   Recalling the logarithmic fit to the numerical integration of the dimensionless Fourier transform, the spectral density reduces to
\begin{equation}
J_1(\omega) \approx \frac{2 \; m_e^2/ e^2}{(2 \pi)^5  c_T \rho}  I_c^2\; \hbar \omega\;a \log[b\;kR]
\end{equation}
which is ohmic ($\omega$ is raised to the first power) but also depends quadratically on the critical current $I_c$ and grows logarithmically with the size of the transmon $kR$ (measured in phonon wavelengths).

\subsection{Thermalization and Dephasing}

In the master equation the (diagonal) population-mixing, or thermalization, vanishes in the thermal balance
\begin{equation}
\frac{\rho_{11}}{\rho_{00}} = \frac{N_{\omega}} {N_{\omega}+1} = e^{-\hbar\omega/k_BT}.
\end{equation}

For the (off-diagonal) coherence $\rho_{01}$ we use $(2N_{\omega}+1)=\coth[\hbar\omega/2k_BT]$ to write the dephasing rate as
\begin{equation}
\Gamma_{10} \equiv \frac{2\pi}{\hbar^2}\frac{1}{2} |\bra{0}\sigma_{\phi}\ket{1}|^2 J_s(\omega) \coth[\frac{\hbar\omega}{2k_BT}].
\end{equation}

As an example, consider the IBM transmon\cite{Rigetti} whose niobium island/pads encompass $\approx 0.5$ mm$^2$. We choose a corresponding model transmon radius of $R=400$ $\mu$m.  Operating at $\omega/2\pi = 4$ GHz and $E_J/E_C=49$, with $I_c=20$ nA, and at low temperatures $k_BT\ll \hbar \omega$, e.g. $10$ mK, we calculate, with $\rho \approx 8570$ kg$/$m$^3$ and $c_T \approx 1600$ m$/$s, a dephasing time of $\Gamma_{10}^{-1} \approx 10$ s.

Operating a similar qubit $10$ times as fast with $10$ times the current, for example by using different junction parameters and thus energy $E_J$, gives a dephasing time of $\approx 7$ ms which is still two orders of magnitude longer than IBM's almost $100$ $\mu$s decoherence time.

\section{Conclusion}

We conclude that transmon phonons are no problem.

\appendix*
\section{Other Polarizations}

For the other two polarizations we make use of a multipole expansion of the periodic step function
\begin{equation}
\frac{\psi}{|\psi|} = \sum_ {m=1,3,...}^{\infty} \frac{4}{m\pi}\sin(m \psi).
\end{equation}
With $\hat{r}\cdot\{\hat{e}_2,\hat{e}_3\}=\cos(\phi-\psi)\{\cos\theta,\sin\theta\}$ we find
\begin{equation}
\tilde{j}_{\vec{k}2,3} = \sum_{m=1,3,...}^{\infty}\frac{4}{m\pi}\sin(m\phi)2\pi \im^{m-1} \frac{J_m[kR \sin\theta]}{\{\tan\theta,1\}}.
\end{equation}

We don't expect longitudinal phonons to have an appreciable effect over many wavelengths.  The in-plane projection of the not-in-plane transverse polarization is similarly longitudinal.  In fact, numerical integrations consistently show these spectral densities to be small fractions of that of the transverse in-plane polarization:
\begin{equation}
\int d^2\Omega \; |\tilde{j}_{\vec{k}1}|^2 \gg \int d^2\Omega \; |\tilde{j}_{\vec{k}2}|^2 \gg \int d^2\Omega \; |\tilde{j}_{\vec{k}3}|^2.
\end{equation}
We thus use only the transverse in-plane polarization in estimating the dephasing rate.

%\begin{acknowledgments}
%\end{acknowledgments}
%\bibliography{transmophonics}

%merlin.mbs 2010-03-15 4.21a (PWD, AO, DPC)

\end{document}